\documentclass[10pt]{iopart}

\usepackage{iopams}  
\usepackage{graphicx}

\begin{document}

\title{Photo-induced modifications of the substrate-adsorbate interaction in K-loaded porous glass}

\author{L. Marmugi$^{1,2}$\footnote{Present address: Department of Physics and Astronomy, University College London, Gower Street, London WC1E 6BT (United Kingdom)}, E. Mariotti$^{1}$, A. Burchianti$^{3}$, S. Veronesi$^{4}$, L. Moi$^{1}$, C. Marinelli$^{1,2}$}
\address{$^1$ CNISM and DSFTA, University of Siena, via Roma 56, 53100 Siena (Italy)}
\address{$^2$ INO-CNR, uos Pisa, via G. Moruzzi 1, 56124 Pisa (Italy)}
\address{$^3$ INO-CNR and LENS, University of Florence, via N. Carrara 1, 50019 Sesto Fiorentino (Italy)}
\address{$^4$ NEST Istituto Nanoscienze - CNR, piazza San Silvestro 12 and Physics Department, University of Pisa, largo B. Pontecorvo 3, 56127 Pisa (Italy)}
\ead{carmela.marinelli@unisi.it}

\date{\today}

\begin{abstract}
The effects of visible and infrared light on potassium atoms embedded in a nanoporous glass matrix are investigated. Photodesorption by visible light enhances the atomic mobility and causes the formation of metallic nanoparticles. Two different populations of metastable clusters with absorption bands in the near-infrared and infrared are grown as a consequence of illumination. Atoms can move between the two groups through sequences of adsorption/desorption events at the pore surface. Irradiation with infrared light, instead, does not significantly enhance the atomic diffusion inside the pores. However, it induces relevant modifications of the substrate, thus changing its interaction with the assembled clusters. Consequently, infrared light alters the dynamics of the system, affecting also the evolution of non-resonant nanoparticles populations, even after the illumination sequence. These results provide new insights on the photo-induced modifications of the substrate-adsorbate interaction in nano-sized confined systems.
\end{abstract}


\vspace{2pc}


\maketitle


\section{Introduction}
A deep understanding of adsorption/desorption and nucleation processes is of great importance for novel engineered materials \cite{xu2013, sukharev2007, singhal2010, tritsaris2013}. Indeed,  the possibility to recognise and modify the underlying mechanisms would allow to manipulate the overall atomic transport dynamics  \cite{brault2009, villalba2010, martins2013}. In this context, because of the increasing  demand of nano-sized devices, the scientific research has focused on the study of  surface interactions at the micro- and nanoscale. In particular, several experiments have proven that light is a powerful tool to trigger and control the atomic adsorption/desorption events and the atomic diffusion under dimensional confinement \cite{slepkov2008, bhagwat2009, srivastava2010, vartanyan2010, perrella2013}.

In this paper, we investigate the influence of visible and infrared (IR) illumination on K atoms confined in a nanoporous glass matrix. In both cases, although light acts on different groups of adsorbed atoms, we observe that the equilibrium among atomic vapour, atomic layers and populations of light-assembled K nanoparticles is changed. As a result, the system moves towards new and different metastable states.

It was recently demonstrated that exposure to visible light increases the K atomic mobility by inducing desorption from the nanopores' inner walls \cite{marmugi2014}. As a consequence, metastable metal nanoparticles self-assemble at surface sites where previously adsorbed atoms act as seeds for clusters. Surface Plasmon Resonances (SPRs) in the near infrared (nIR) were attributed to light-grown metastable spheroids with average radius of 2 nm.

Here, we identify a further light-grown population of K aggregates with absorption band in the IR region. The case of exposure to IR light, which is resonant with this last nanoparticles' distribution, is then explored. We found that, contrary to the visible case,  infrared light does not produce any substantial enhancement of the bulk atomic diffusion. Nevertheless, it locally induces, even at low intensity, structural changes of  the substrate lying at the pores' surface and formed by adatoms, wide distributions of undistinguished nanoparticles and the residuals of evaporated nanostructures. Therefore, the interaction between K atoms and the substrate is modified.

\section{Experimental apparatus}
Light-enhanced diffusion and nanoparticles formation are observed inside the randomly oriented pores of a porous glass (PG) plate (18$\times$15$\times$1 mm$^{3}$), manufactured by VitraBio GmbH (see left inset in Fig. \ref{fig:efficiency}). The pores have an average diameter of 20 nm. The internal surface of the PG matrix is about 31 m$^{2}$, with a 55$\%$ free volume with respect to the sample footprint. The PG sample is fixed inside a sealed Pyrex cell, where it is continuously exposed to the K vapour supplied by a solid reservoir at room temperature (see right inset in Fig. \ref{fig:efficiency}). Atoms progressively enter the pores, where they perform a random free-flight, which can be interrupted by adsorption at surface sites along the inner walls \cite{villalba2010}. 

Further details on the experimental apparatus are given in Ref. \cite{marmugi2014}. Here, we only recall that the K vapour density in the cell free volume is measured by detecting the laser-induced atomic fluorescence from the D$_{1}$ line transition at 770.1 nm. The sample transparency is monitored by two free-running, continuous wave (CW) laser diodes, tuned at 780 nm and 1.46 $\mu$m, respectively; these probes (Transmission Probe Beams, TPBs) are resonant to the main absorption bands produced by the nanoparticles inside the pores' network. In order to prevent perturbations of the system due to thermal effects, the power of the TPBs is constantly kept below 400 $\mu$W and below 100 $\mu$W for the 780 and the 1.46 $\mu$m TPBs respectively. The probe beams have a waist of 5 mm at the surface of the sample and they are overlapped in order to probe the same PG region. Moreover, because of the high light-scattering efficiency of the porous glass, photons are diffused within the sample and the intensity is further lowered.

Photodesorption is induced either by a frequency doubled CW Nd:YAG laser (532 nm) or by a 405 nm free running laser diode. An opposite regime, where the atomic diffusion enhancement is negligible, is instead obtained by means of an 1.55 $\mu$m free-running laser diode.

The absorbance spectra of the sample, before and after external illumination, are acquired with a commercial spectrophotometer (Varian Cary 500). Each spectrum is referenced to the equilibrium absorbance:

\begin{eqnarray}
Abs_{ref} (\lambda, t)   & =  Abs(\lambda, t)-Abs_{0}(\lambda)  \nonumber \\
{} & = - \log \left( \frac{T(\lambda, t)}{I} \right) + \log \left( \frac{T_{0}(\lambda)}{I}\right)  \mbox{ ,}\label{eqn:absir}
\end{eqnarray}

where $I$ is the incident light intensity, $T_{0}(\lambda)$ is the light intensity transmitted through the porous sample at the equilibrium in the dark and $T(\lambda, t)$ is instead the light intensity transmitted through the PG after an illumination of duration $t$.

\section{Experimental results}
\subsection{Photodesorption processes}
The photodesorption  efficiency is measured by illuminating the PG sample with laser light ranging from the visible to the infrared. Two parameters are introduced, measuring respectively the maximum relative variation of the K vapour density in the cell free volume, $n$, upon illumination ($\delta_{max}$) and its growth rate at the beginning of the exposure to the laser radiation ($R$):

\begin{eqnarray}
\delta_{max} & \equiv &\frac{n(t)|_{max}-n_{0}}{n_{0}} \label{eqn:delta} \mbox{ ,}\\
R & \equiv &\frac{1}{n_{0}}\frac{dn(t)}{dt}\Big|_{0^{+}} \label{eqn:rate} \mbox{ .}
\end{eqnarray}

In the previous definitions, $n_{0}\approx 4.6\times 10^{8}$ cm$^{-3}$ is the equilibrium value of the vapour density at room temperature in absence of  illumination. While $R$ (Eq. \ref{eqn:rate}) is more sensitive to the atoms desorbed from sites closer to the external surface of the PG matrix,  $\delta_{max}$ gives an integral measure of the desorption, including both the contributions from more superficial sites and from the innermost regions of the pores.

\begin{figure}[h]
\begin{center}
\includegraphics[width=\linewidth]{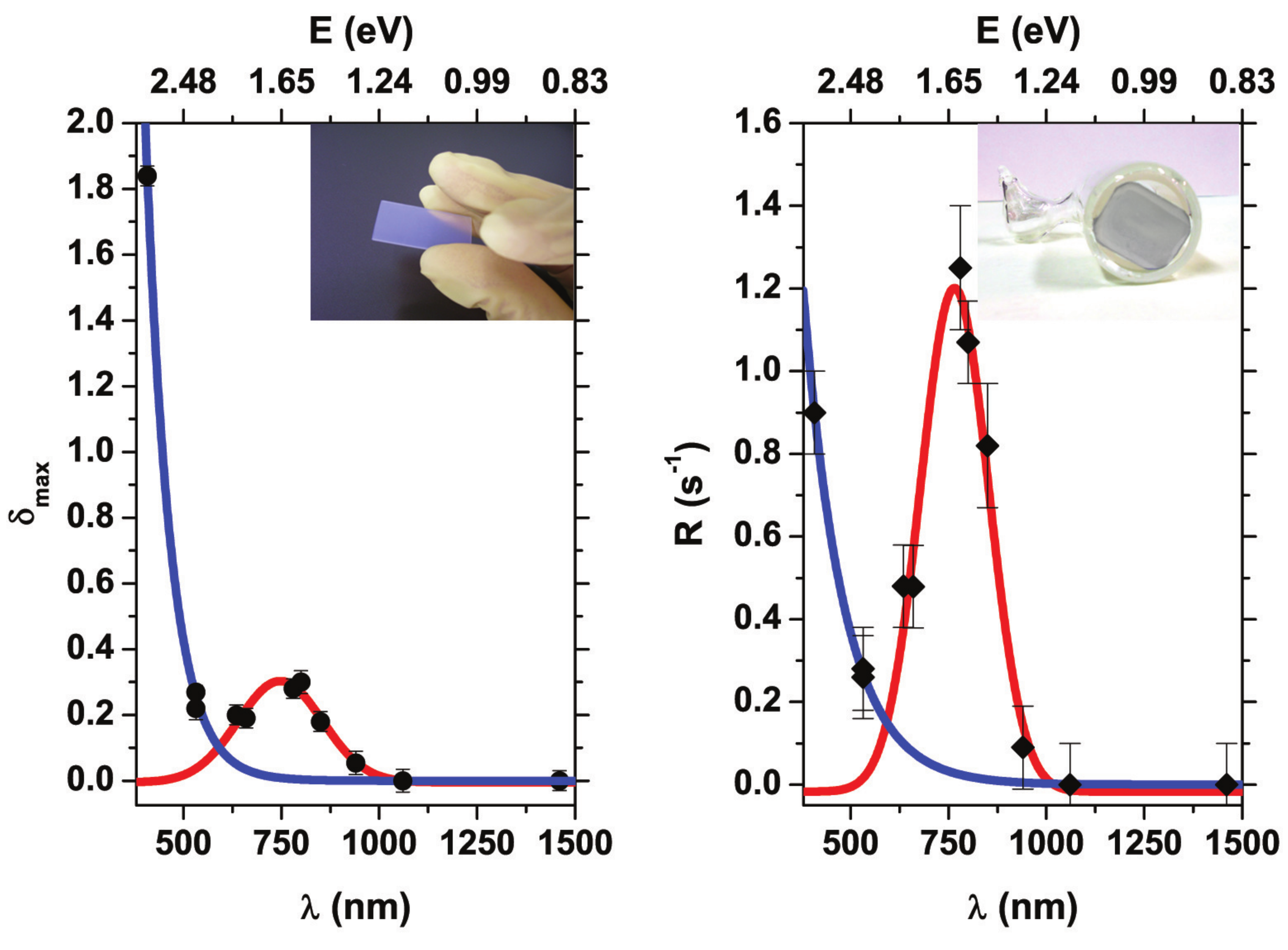}
\caption{$\delta_{max}$ and $R$ as a function of the desorbing light wavelength at the fixed intensity of 5 mW/cm$^{2}$. Blue curve: LIAD non-resonant exponential contribution, given by $f(\lambda)=A \exp \left( -\frac{\lambda}{\theta} \right)$. Red curve: SPID resonant Gaussian contribution, given by $g(\lambda)=\frac{B}{w \sqrt{\pi/2}} \exp \left[ -2 \left( \frac{\lambda-\lambda_{c}}{w} \right)^{2} \right]$. Details about the fit results are reported in Tab. \ref{tab:fits}. Left inset: photograph of the commercial PG sample. Right inset: photograph of the sample inserted in the Pyrex cell. The K solid reservoir is visible on the left side.}\label{fig:efficiency}
\end{center}
\end{figure}

\begin{table}[h]
\footnotesize\rm
\begin{tabular*}{\columnwidth}{@{}l*{15}{@{\extracolsep{0pt plus12pt}}l}}
\hline
\textbf{Parameter} & \textbf{Process} & \textbf{Fit parameters} & \textbf{Goodness of fit} \\
\hline
$\delta_{max}$ & LIAD & $A=1185 \pm 242$ & Adj. R square=0.99\\
 &  & $\theta=62 \pm 2$ nm & Red. $\chi^{2}$=0.77\\
 \hline
 & SPID & $B=83 \pm 19$ nm & Adj. R-square=0.92\\
 &  & $\lambda_{c}=747 \pm 9$ nm & Red. $\chi^{2}$=0.87\\
 &  & $w=216 \pm 37$ nm & \\
\hline
\hline
$R$ & LIAD & $A=47 \pm 5$ s$^{-1}$  & Adj. R-square=0.99 \\
 &  & $\theta=103 \pm 3$ nm s$^{-1}$ & Red. $\chi^{2}$=0.007\\
 \hline
 & SPID & $B=265 \pm 45$ nm s & Adj. R-square=0.95 \\
 &  & $\lambda_{c}=765 \pm 10$ nm & Red. $\chi^{2}$=0.91\\
 &  & $w=174 \pm 23$ nm & \\
\hline
\end{tabular*}
\caption{Fit parameters for Fig. \ref{fig:efficiency}.}\label{tab:fits}
\end{table}

Results shown in Fig. \ref{fig:efficiency} indicate the existence of two different desorption processes: the non-resonant, exponential contribution from the Light-Induced Atomic Desorption (LIAD) and the resonant one, centred at 747$\pm$9 nm (1.66$\pm$0.02 eV), produced by the direct excitation of surface plasmons and subsequent evaporation of light-grown K nanostructures (Surface Plasmon Induced Desorption, SPID) \cite{burchiantiepjd}. Comparison with previous results and cross-check among different samples ruled out possible correlations of these effects with production methods, or handling of the porous glass before its installation in the Pyrex cell.

The two processes act on different groups of atoms: the ones lying along the pores' surface in the first case and the ones assembled in resonant nanoparticles in the second. In any case, they both force atoms to resume their free-flight along the pores: after a sequence of adsorption and desorption events, a fraction of the atoms reaches the vapour phase and increases its density \cite{villalba2010}. 

The measured resonance frequency is consistent with the predictions for the SPR of K nanoparticles embedded in the PG matrix:

\begin{equation}
\hbar \omega_{SPR}=\frac{\hbar \omega_{p}}{\sqrt{1+2\varepsilon_{eff}}}=1.65 \mbox{ eV} \mbox{ ,}
\end{equation}

where $\hbar \omega_{p}=3.72$ eV is the energy of the K bulk plasma oscillation \cite{raether1980} and $\varepsilon_{eff}=2.04$ is the effective dielectric constant of the porous glass \cite{burchiantiepjd}.

From Fig. \ref{fig:efficiency}, it is evident that: i. if only the more superficial contribution is taken into consideration, photodesorption due to SPID can be more relevant than the LIAD one; ii. regardless the mechanisms involved, no significant variations of the vapour density are observed for $\lambda\geq 1100$ nm. It is also worth noting that the enhancement of atomic diffusion through the pores is more effective in the case of LIAD than in the case of SPID. In particular, this indicates also that the majority of K clusters tends to accumulate in the proximity of the pores' aperture, where the largest flux from and to the vapour phase is expected in equilibrium conditions. Visible light, instead, as a consequence of the higher atomic mobility, provides a more relevant contribution of diffusion from the bulk to the vapour density. Consistently, SPID is prevalently induced in the pores' region closer to the PG's external boundary, whereas LIAD produces a significant wavelength-dependent diffusion from the inner volume \cite{marmugi2014}.

\subsection{Visible illumination: atomic diffusion enhancement and role of the substrate}\label{sec:visible}
\begin{figure}[htb]
\begin{center}
\includegraphics[width=\linewidth]{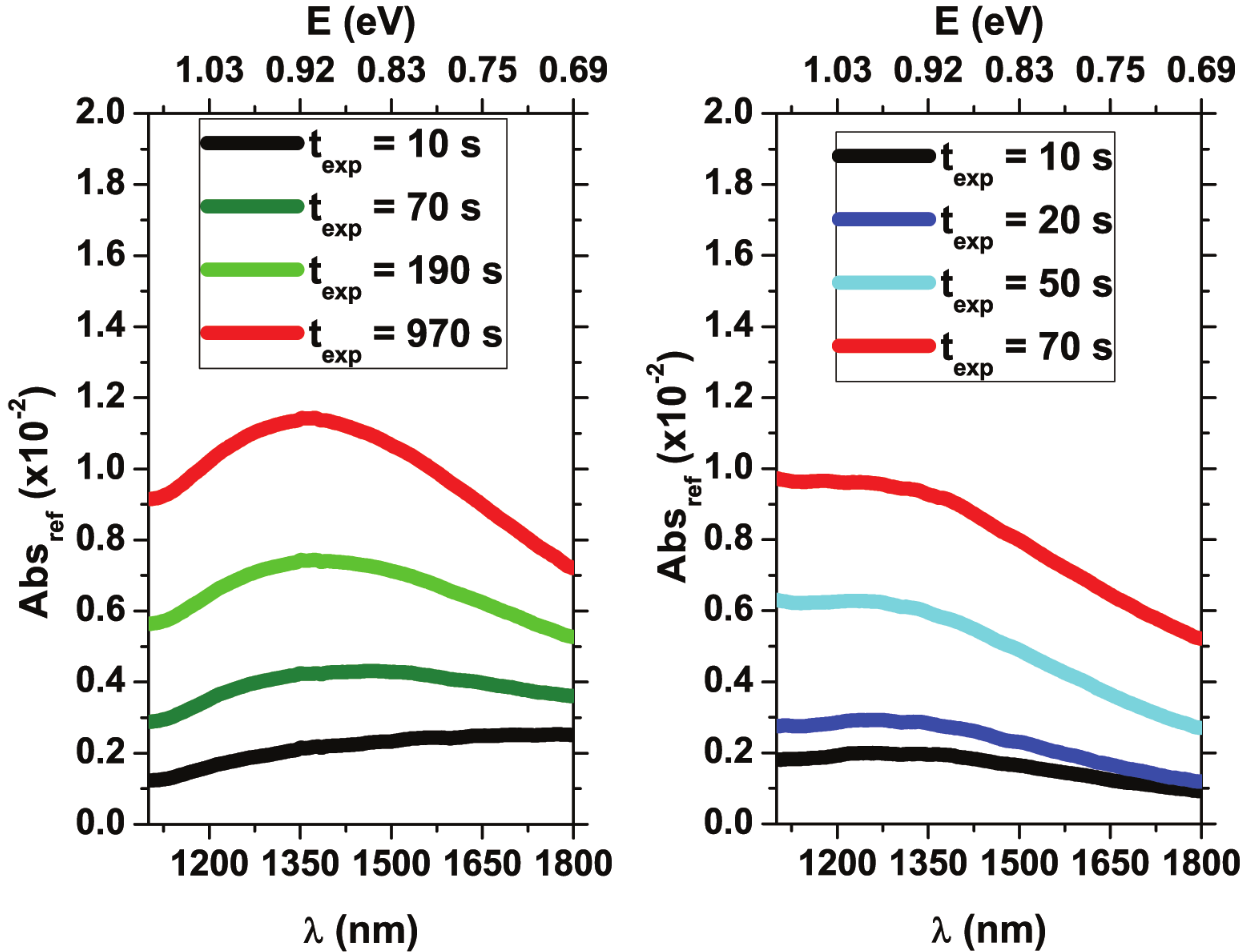}
\caption{Infrared absorbance of the PG sample during an illumination cycle with visible light. \textbf{Left:} 350 mW/cm$^{2}$ at 532 nm.  \textbf{Right:} 5 mW/cm$^{2}$ at 405 nm.}\label{fig:absgreen}
\end{center}
\end{figure}

In Fig. \ref{fig:absgreen} the PG absorbance in the IR region is shown after different cycles of visible illumination.

Visible light causes a progressive increase of the sample absorbance. A two-maxima peak in the near infrared (nIR) band is produced by the SPRs at the boundaries of K spheroids with 2 nm mean radius. This phenomenon was thoroughly analysed in Ref. \cite{marmugi2014} and therefore is not shown here. The K spheroids are the most relevant class of nanoparticles self-assembled as a consequence of illumination, but other more irregular shapes of nanoclusters are present, either as deformed structures or as intermediate stages of the NPs' formation. Nevertheless, their contribution to the nIR absorbance peaks is small compared to the spheroids' one. In addition, as demonstrated by Fig. \ref{fig:absgreen}, a broader absorbance peak appears in the infrared (IR) spectral interval between 1.1 $\mu$m and 1.8 $\mu$m. The peak has a Gaussian shape and its height increases with the exposure time to the visible desorbing light. It is evident from Fig. \ref{fig:absgreen} that shorter visible wavelengths are more efficient in producing NPs with IR absorption. The amount of such nanoparticles, however, is small compared to the ones with definite size and geometry observed in the nIR band.

No signatures of internal spectral structures can be found.  Nevertheless, as suggested by the direct comparison between the two graphs in Fig. \ref{fig:absgreen}, the shape of the resonance exhibits features that depend on the desorbing wavelength. In fact, the shorter wavelength part of the resonance ($\lambda \leq 1300$ nm) increases faster than the longer one ($\lambda \geq 1600$ nm). This produces a deformation of the peak, with an increasing asymmetry during illumination, which is more pronounced in the case of blue illumination (right panel of Fig. \ref{fig:absgreen}). Such result indicates a direct connection between the shape of the IR absorbance peak and the light-induced increase of atomic mobility.

\begin{figure}[htbp]
\begin{center}
\includegraphics[width=\linewidth]{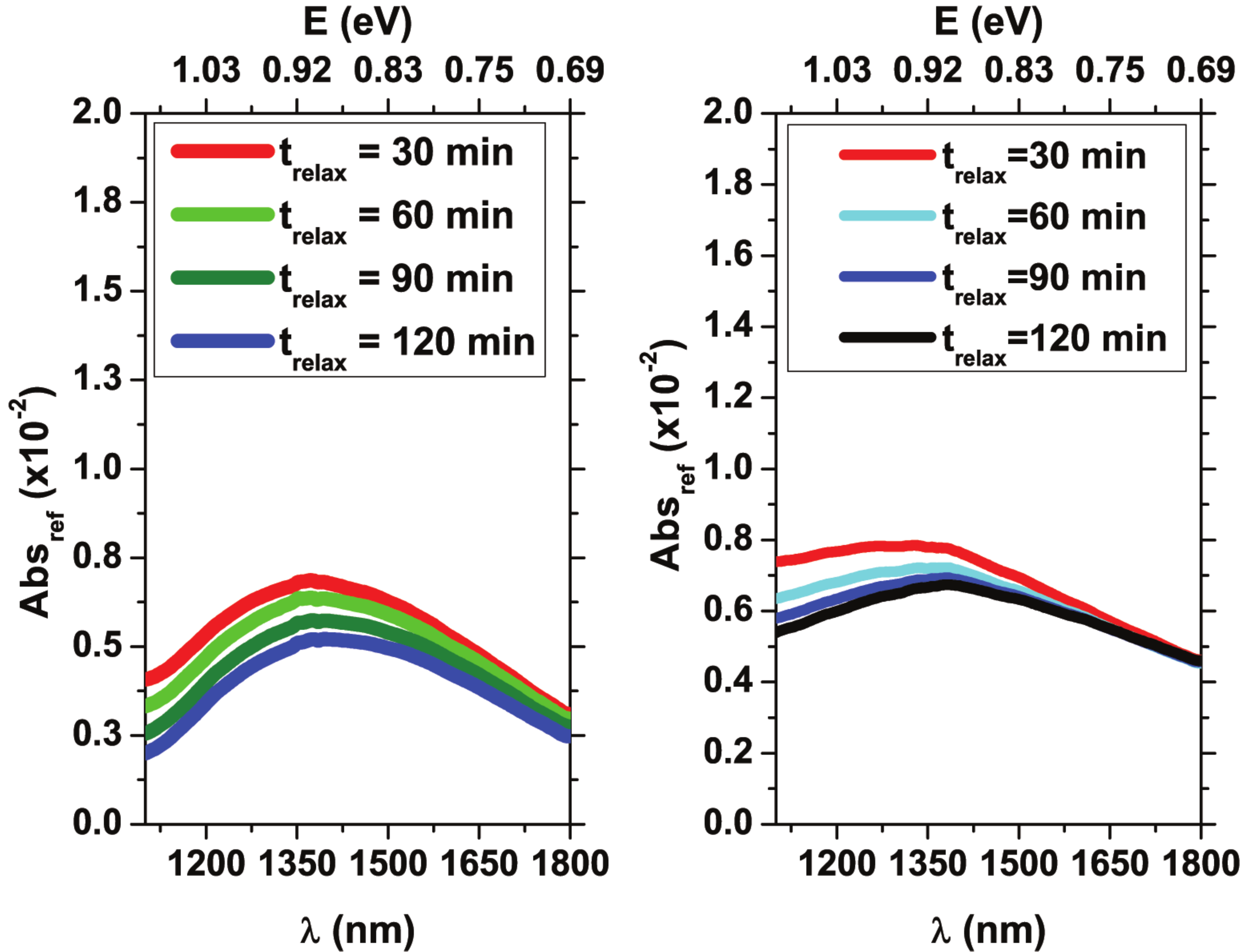}
\caption{IR absorbance of the PG sample during relaxation in the dark. \textbf{Left}: after exposure to 350 mW/cm$^{2}$ at 532 nm for 970 s. \textbf{Right}: after exposure to 5 mW/cm$^{2}$ at 405 nm for 70 s.}\label{fig:irrelax}
\end{center}
\end{figure}

In the dark, the asymmetry between the left and the right ``tails'' of the resonance is progressively reduced (see Fig. \ref{fig:irrelax}): the shorter wavelength part of the spectrum in fact decreases faster than the other. The behaviour in absence of light suggests the existence of intrinsic spontaneous decay mechanisms independent from the previous illumination.

\begin{figure}[htb]
\begin{center}
\includegraphics[width=\linewidth]{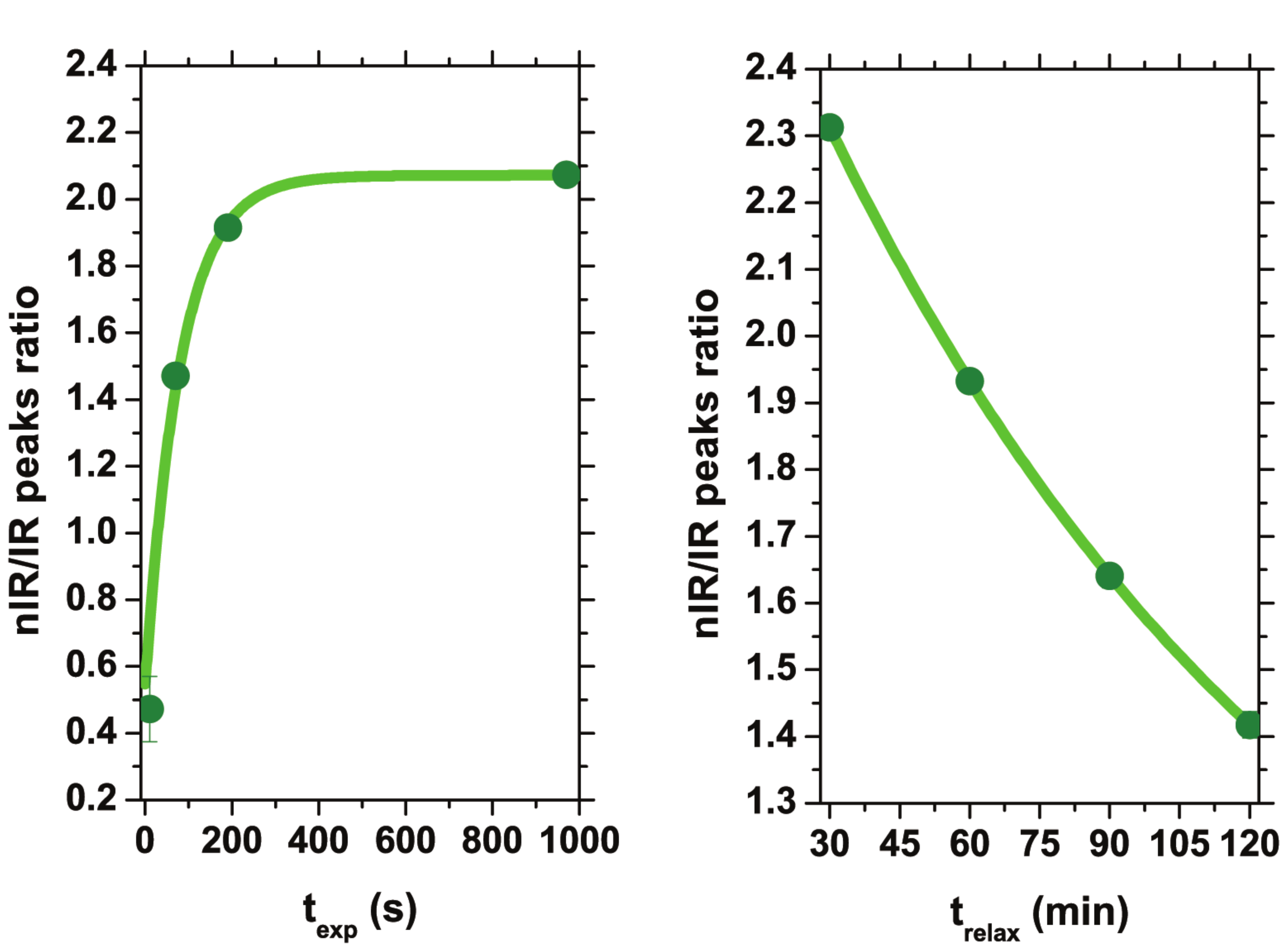}
\caption{Relative variation of the nIR and IR absorbance peaks. \textbf{Left}: illumination with 350 mW/cm$^{2}$ at 532 nm. \textbf{Right}: relaxation in the dark. Continuous lines are exponential fits of the form $f(t)=A \exp \left( -\frac{t}{\theta} \right)+f_{0}$. The fitting time constants are $\theta_{exp}$=81$\pm$11 s (Left panel) and $\theta_{relax}$=113$\pm$1 min (Right panel), respectively.}\label{fig:relative}
\end{center}
\end{figure}

In addition, the infrared peak decays more slowly than spheroids' ones: the IR absorbing aggregates hence are intrinsically more stable than the nIR ones. As a demonstration, in Fig. \ref{fig:relative}, the relative increase of the population of the 2 nm spheroids according to Ref. \cite{marmugi2014} is compared to the one measured for the IR peak, in the case of green illumination and following relaxation. It is evident that, both in presence of desorbing visible light and in the dark, the evolution of the infrared peak is produced in time scales different from the spheroids' one.

Consistently with previous results, the $\lambda$-dependent evolution of the IR absorbance peak is  explained by the presence of metastable K nanoparticles, self-assembled along the nanopores' walls as a consequence of light-enhanced atomic mobility and nucleation at surface defects. In particular, a higher NPs' instability is obtained with a higher light-enhanced atomic mobility.

The experimental evidences discussed so far indicate that the IR peak is produced by a different, independent population with respect to the spheroids' one. Indeed, wide, structureless resonance bands are typically observed whenever a broad distribution of clusters' radii is present: the size spread masks the single SPR contributions, thus smearing the resonances peaks over a broad band \cite{kreibig1995}. As a consequence, no preferred NPs' sizes or geometries are present, unlike the case of the nIR band. The IR peak can be thus explained by wide K NPs, with average radius larger than the nIR spheroids and a random distribution of aspect ratios.

\begin{figure}[htbp]
\begin{center}
\includegraphics[width=\linewidth]{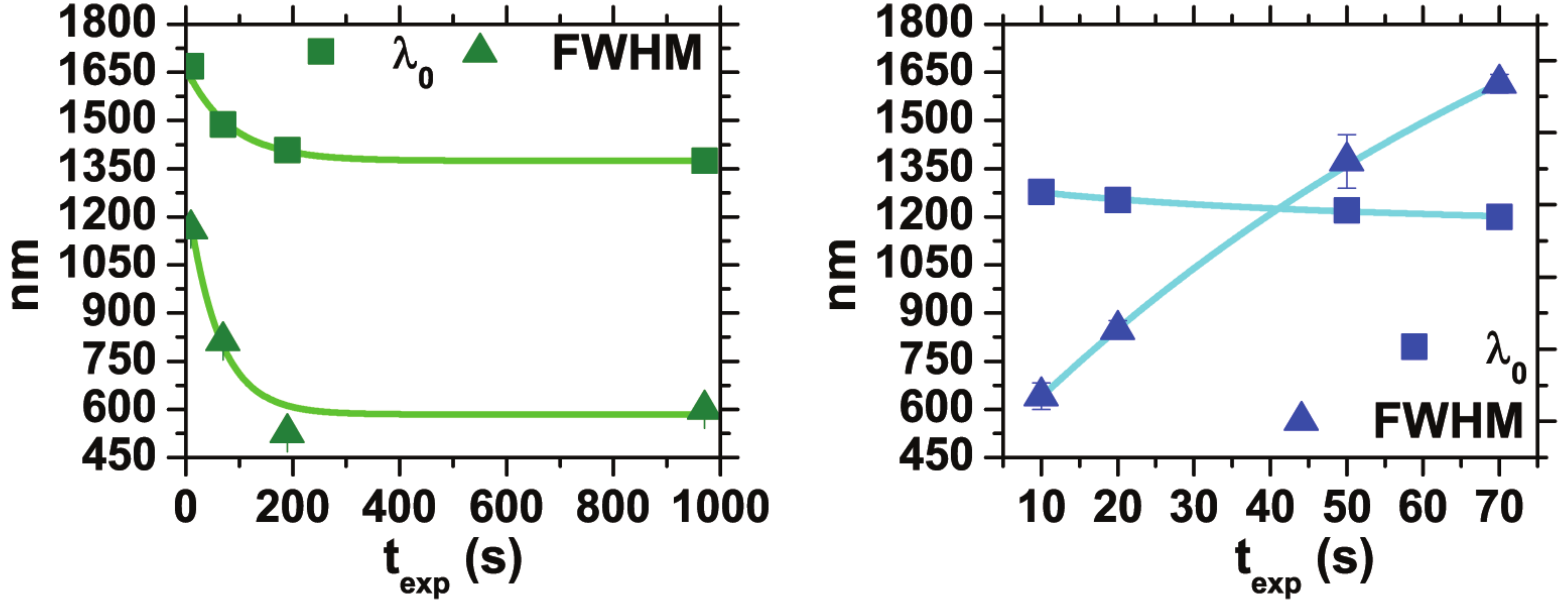}
\caption{\textbf{Left:} IR absorbance peak position ($\lambda_{c}$) and FWHM as a function of exposure time to 350 mW/cm$^{2}$ at 532 nm. \textbf{Right:} $\lambda_{c}$ and FWHM as a function of exposure time to 5 mW/cm$^{2}$ at 405 nm. Continuous lines are exponential functions shown as a guide to the eye.}\label{fig:fits}
\end{center}
\end{figure}

As demonstrated by Fig. \ref{fig:fits}, the random accumulation of broader structures continues as the illumination proceeds. The progressive change of the peak's position and width reveals a rearrangement of NPs size and geometry. In fact, the asymmetry between the left and the right ``tails'' of the resonance peak produced during illumination forces $\lambda_{c}$ to shift towards shorter wavelengths. In terms of NPs, this is caused by the relative increment of smaller structures' number, as a consequence of the atomic mobility enhancement and shaping effects induced by the desorbing light. Consistently, the effects are more pronounced in the case of the blue illumination, when a more relevant increase of the average atomic mobility is produced.

In fact, the blue desorbing light removes neighbouring atoms more effectively than green, thus restraining the coalescence of large structures. At the same time, it shapes more efficiently the growing clusters by moving the less bound, external atoms, thus progressively reducing the aggregate transverse dimensions. In summary, the larger atomic mobility imposed by the light-induced atomic desorption enhances the deformation of the IR resonance peak during illumination, thus suggesting the progressive self-assembly of nanostructures with a broader distribution of size and aspect ratio, but with a smaller average radius.

This result confirms the importance of the surface diffusion and of the interaction with an evolving substrate, both influenced by light. Light speeds up atomic mobility inside the pores, thus enhancing the probability of cluster formation at surface defects. However, during nucleation of nanostructures, light-enhanced diffusion continues to affect atoms captured by the NPs, by modifying the surrounding environment and the shape of the growing structures. As a consequence, also in the case of the IR population, visible light acts as a maker and as a shaper for K nanostructures.

\begin{figure}[htb]
\begin{center}
\includegraphics[width=\linewidth]{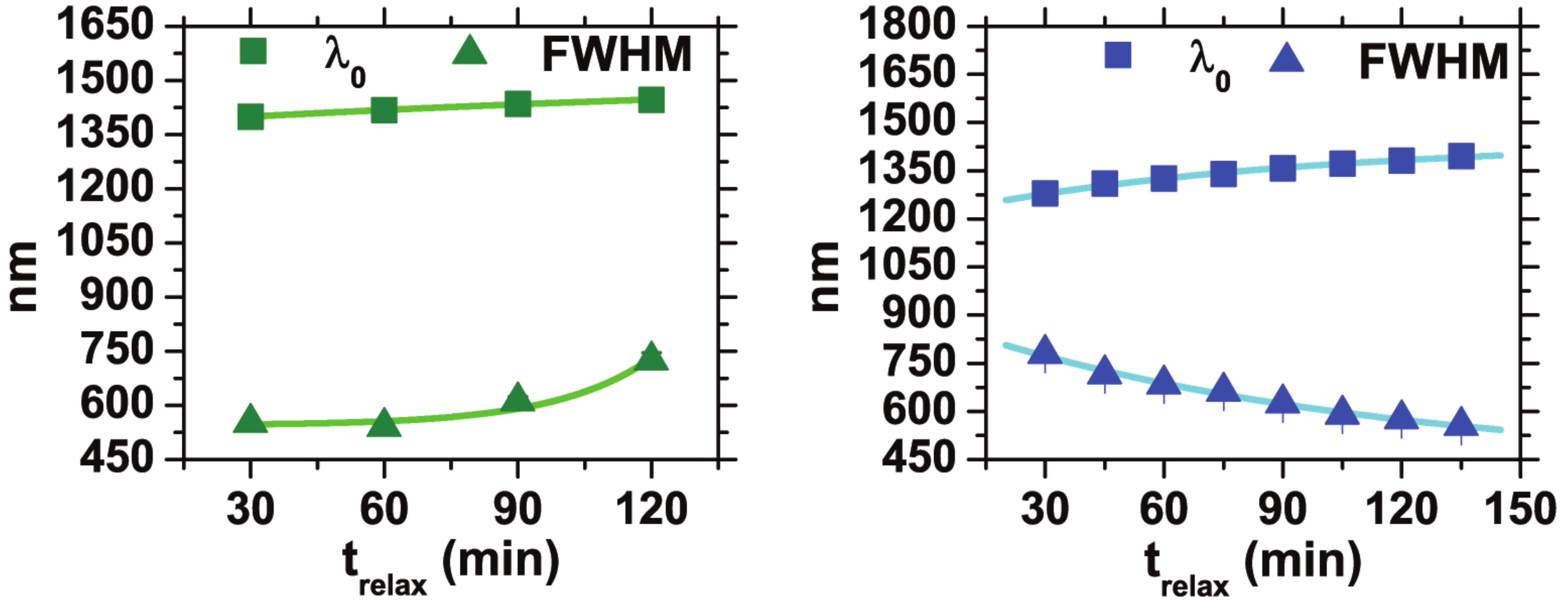}
\caption{\textbf{Left:} $\lambda_{c}$ and FWHM during relaxation in the dark after exposure to 350 mW/cm$^{2}$ at 532 nm for 970 s. \textbf{Right:} $\lambda_{c}$ and FWHM during relaxation in the dark after 70 s exposure to 5 mW/cm$^{2}$ at 405 nm. Continuous lines are exponential functions shown as a guide to the eye.}\label{fig:fitsrelax}
\end{center}
\end{figure}

The behaviour of $\lambda_{c}$ and FWHM during relaxation in the dark (Fig. \ref{fig:fitsrelax}) is consistently opposite to the one observed during illumination: as a result of the spontaneous processes causing the relaxation, the system quickly washes away the modifications imposed by the desorbing light.

The evolution after green and blue illuminations is caused by the combined effect of the decay of light-grown structures, followed by further nucleations and coalescence due to the atoms released inside the nanopores. A cross-talk mechanism during relaxation in the dark can be inferred: as soon as the unstable light-grown nIR spheroids evaporate in the dark, more and more atoms resume their random free-flight inside the pores. At the same time, due to the residuals of NPs' evaporation, the number of surface defects is enlarged. As a consequence, new broader nanostructures are spontaneously assembled and hence the IR absorbing peak gets larger and broader, given the absence of the shaping action of the desorbing light \cite{marmugi2014}. This process can be described as a spontaneous transfer of population from the spheroids' distribution to the IR absorbing one. This phenomenon is however mediated by atomic diffusion at the surface, as it is in all the interface processes \cite{brault2009, lenzi2010, sapora2014}.

As a result, also the FWHMs of the resonance become again comparable in the two cases (Fig. \ref{fig:fitsrelax}): the spontaneous evaporation of the metastable IR nanoclusters re-creates an equilibrium distribution of aggregates regardless the previous conditions. In this context, consistently with the behaviour of the nIR spheroids, the aggregates self-assembled in presence of a larger atomic mobility exhibit larger instability.

\subsubsection{Evolution of NPs' populations driven by visible light \\}
In this section, we investigate the dynamics of the nIR and IR absorbing nanoparticles by recording the TPBs at 780 nm and at 1.46 $\mu$m. K clusters' growth is induced by uniformly illuminating the sample with 40 mW/cm$^{2}$ at 532 nm. In order to study the interplay between atomic diffusion and NPs growth, we also monitor the atomic vapour density in the cell volume during and after illumination.

\begin{figure}[htb]
\begin{center}
\includegraphics[width=\linewidth]{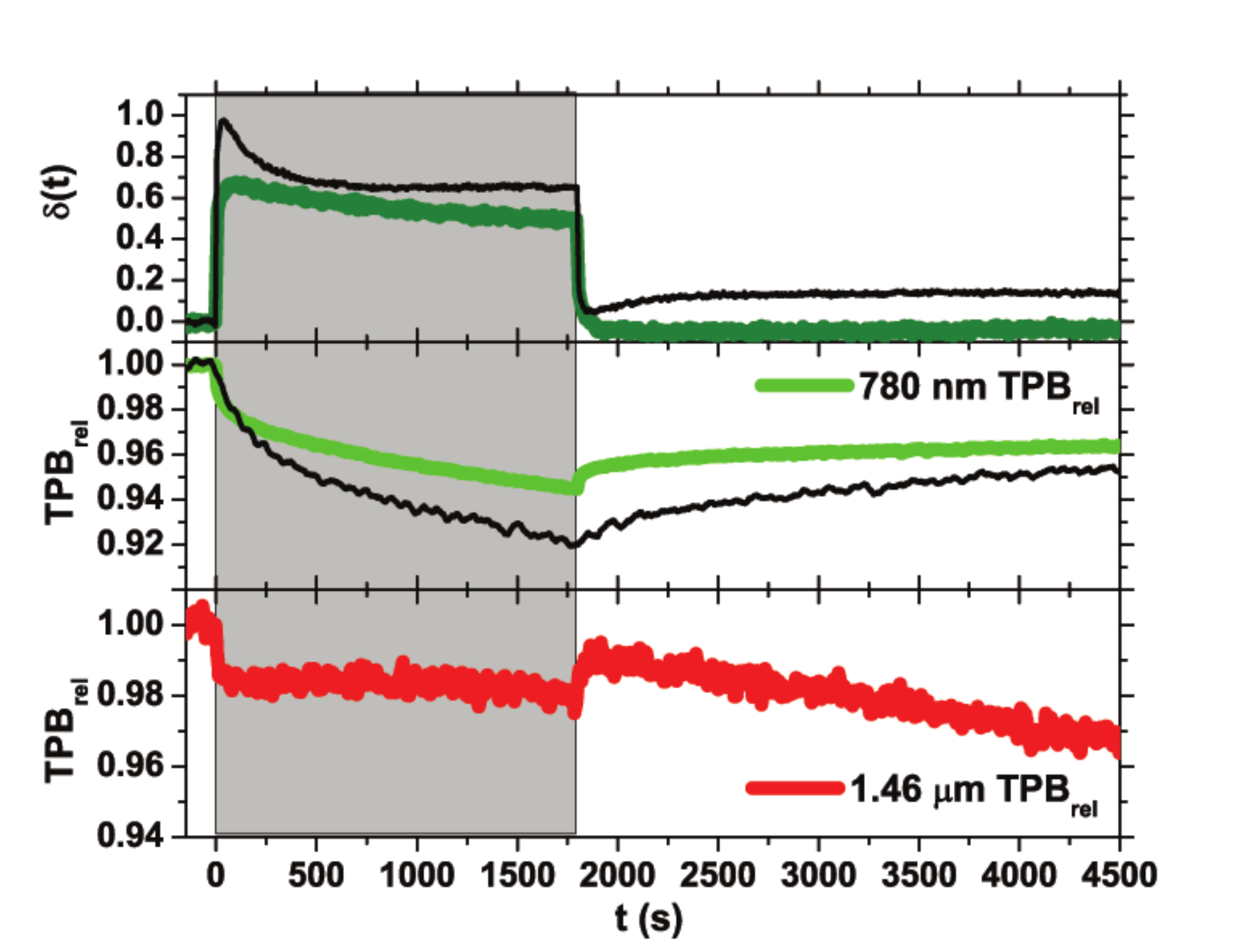}
\caption{Relative variation of the K vapour density $\delta$ (\textbf{upper panel}) and TBPs at 780 nm (\textbf{middle panel}) and 1.46 $\mu$m (\textbf{bottom panel}) as a function of time during uniform illumination with 40 mW/cm$^{2}$ at 532 nm. Desorbing light is switched on at t=0 s and off at t=1800 s, as indicated by the shaded area. The black thin curves are the corresponding traces obtained without the 1.46 $\mu$m TPB.}\label{fig:irdynamics}
\end{center}
\end{figure}

In Fig. \ref{fig:irdynamics},  $\delta(t)$  and the nIR and IR TPBs' signals are shown for an illumination lasting 1800 s.

As soon as the desorbing light hits the sample, the atomic diffusion through the pores increases: some atoms are ejected out of the glass matrix, while others are trapped at the pore surface and form metallic aggregates. As a consequence, $\delta(t)$ increases, reaches its maximum value and, then, starts to drop. In fact, due to the sample depletion, the desorbing flux does not compensate anymore the atomic losses at the pore walls and the adsorption at the solid reservoir. At the same time, the two TPBs decrease, because of the formation of the 2 nm spheroids and of the broader aggregates. However, they exhibit a remarkably different dynamical behaviour. The largest increase of the IR absorbing NPs population is observed immediately after the green laser light is switched on. Later on, the  IR cluster formation rate drops down. On the contrary, the number of the nIR absorbing NPs progressively rises during the illumination.

This can be explained by taking into account the progressive shaping effect produced by the desorbing light: eventually, a balance among the new formations, population transfer and intrinsic instability of the formed clusters occurs. Moreover, a transfer of atoms from the spheroids to the IR aggregates can be inferred from the transmission signals during the relaxation in the dark. As the desorbing light is turned off, in fact, the highly unstable spheroids start to evaporate. Atoms are thus released into the glass nanocavities  and are then re-adsorbed at the pore surface, thus forming, without the shaping effect induced by the green light, broader and almost indistinct metallic aggregates. Therefore, in the dark, while the transmission at 780 nm increases, the one at 1.46 micron further decreases. After several hours, nevertheless, we observe that the initial equilibrium condition is restored.

Unexpectedly, we found that even low-power IR laser irradiation, as the one provided by the TPB at 1.46 $\mu$m, alters the evolution of the non-resonant NPs' population, as well as the photo-desorption efficiency. This emerges by comparison of the green traces of Fig. \ref{fig:irdynamics} with the black thin ones, which are obtained in the same conditions but without the IR probe beam. It is important to stress that no resonance is possible with the 2 nm spheroids: therefore, the 780 nm TPB is modified only as a consequence of indirect processes mediated by the substrate. Our hypothesis is that the 1.46 $\mu$m beam, despite its low intensity, modifies the substrate structure. Thus, the presence of  IR light influences the adsorption/desorption and nucleation processes inside the matrix. In particular, both the atomic diffusion and the formation of spheroids are hindered. This is confirmed by the decrease of the number of atoms photo-ejected in the cell volume and by the smaller variation of the 780 nm transmission signal.  In the next section, we will provide an explanation for this effect.

\subsection{Infrared illumination: negligible diffusion and enhancement of the substrate influence}
\begin{figure}[htb]
\begin{center}
\includegraphics[width=\linewidth]{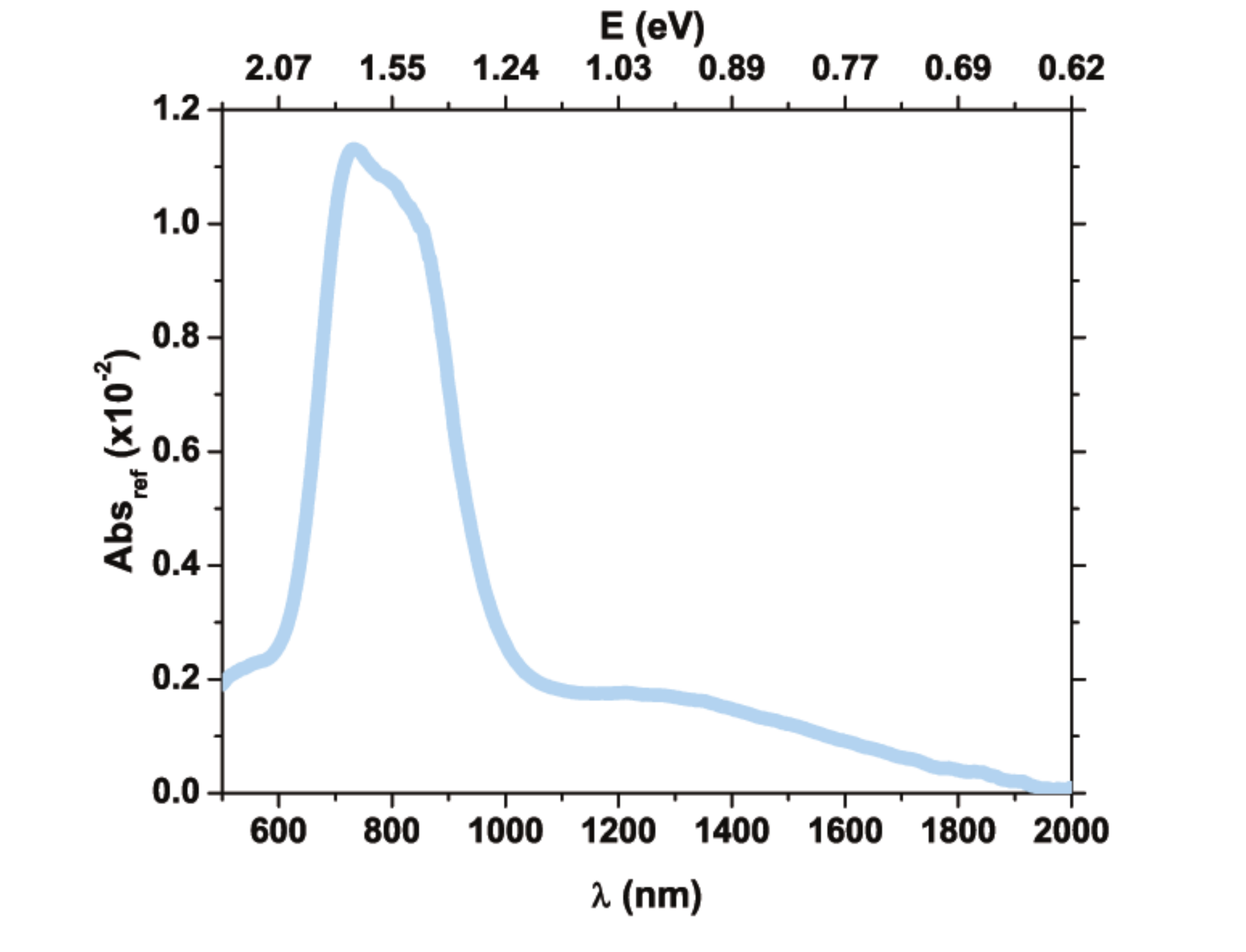}
\caption{Absorbance of the PG sample after an illumination of 10 s with 5 mW/cm$^{2}$ at 405 nm.}\label{fig:absIR}
\end{center}
\end{figure}

In this section, we investigate the effects produced by intense illumination with IR radiation at 1.55 $\mu$m on the substrate-adsorbate complex. For light at this wavelength, the LIAD effect is not observed and, at the equilibrium in the dark, also the SPID effect does not produce significant contributions (Fig. \ref{fig:efficiency}). Therefore, a non-negligible amount of  IR absorbing nanoparticles must be created. For this purpose, the PG sample is uniformly illuminated with 5 mW/cm$^{2}$ at 405 nm for 10 s.  Two well-separated plasmon bands are thus produced (Fig. \ref{fig:absIR}): one centred around 750 nm, due to 2-nm spheroids, and a second one located at 1.28 $\mu$m, produced by the broader aggregates with a non-well-defined shape. The double structure in the first band is due to the formation of prolate and oblate nanoparticles with maximum SPRs at 730 nm and 830 nm, respectively \cite{marmugi2014}.

\begin{figure}[htb]
\begin{center}
\includegraphics[width=\linewidth]{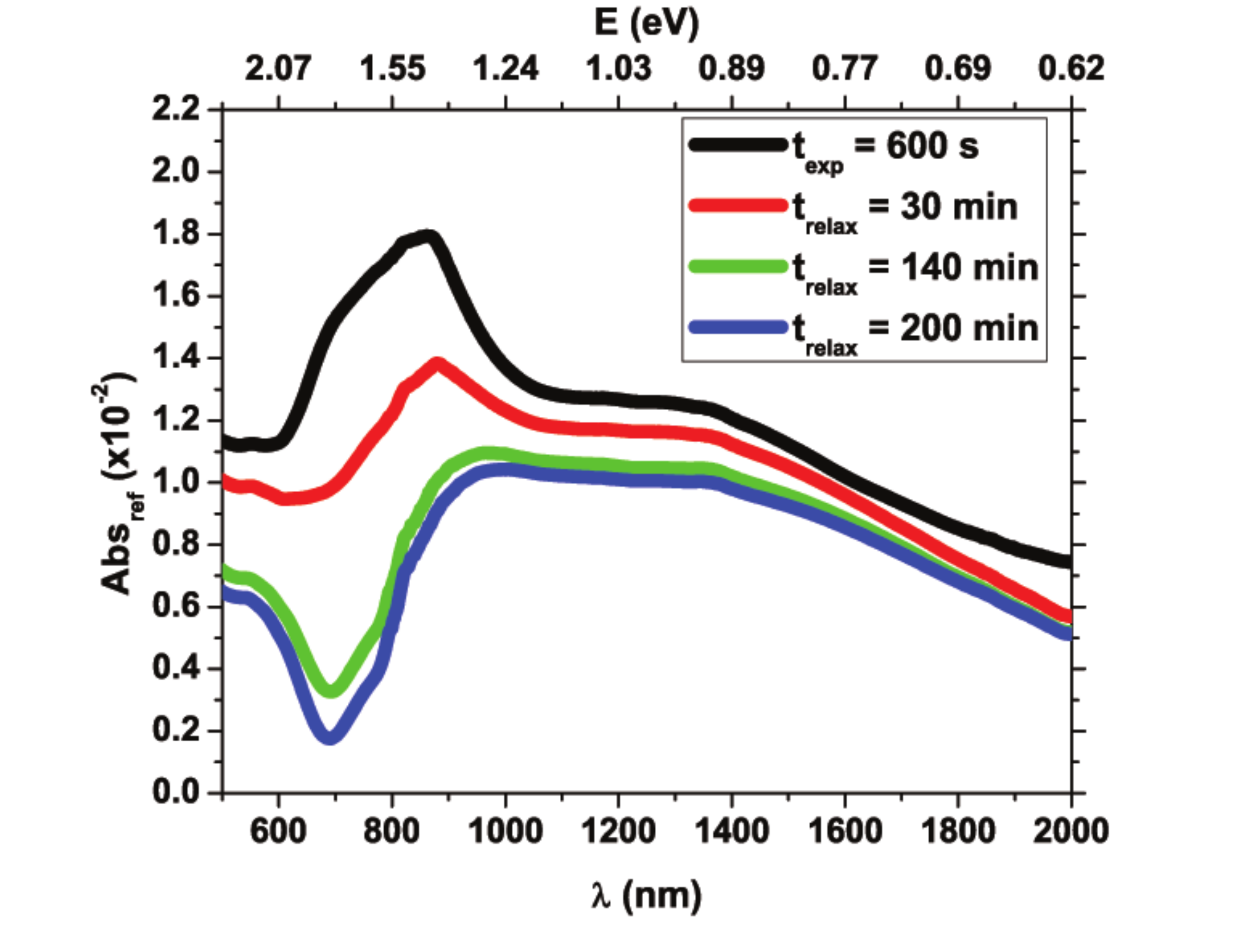}
\caption{Absorbance of the PG sample during relaxation in the dark, after an illumination cycle with 22 mW/cm$^{2}$ at 1.55 $\mu$m.}\label{fig:IRvsblue}
\end{center}
\end{figure}

After exposure to the blue light, the sample is uniformly illuminated with 22 mW/cm$^{2}$ at 1.55 $\mu$m for 600 s. In Fig. \ref{fig:IRvsblue}, the absorbance immediately after IR illumination and during relaxation in the dark is shown. Under these conditions, IR light should force the direct evaporation of the resonant NP population.

However, since the atomic mobility through the pores is not substantially enhanced, the evaporated atoms are likely trapped again at neighbouring sites, where they contribute to form non-resonant broader structures, or metallic islands.  Such aggregates exhibit a very broad absorption band, ranging from the visible to the far-IR regions. As a consequence, no holes appear in the IR spectrum but, on the contrary, the sample absorbance increases and the peak is flattened in its ``red'' tail. Furthermore, these metallic islands with a large surface contact area, result intrinsically more stable: therefore, atoms trapped within such structures have a low probability of further desorption.

The absorption spectra shown in Fig. \ref{fig:IRvsblue} not only reveal the overall increase of broad IR absorbing structures, but also the change of the nIR peak. It was previously observed that the K spheroids' decay, in absence of external illumination, intrinsically depends on their shape; in particular, prolates are structurally less stable and thus evaporate faster than oblates  \cite{marmugi2014}. Here, we find that the geometry-related instability of the spheroids is further enhanced by exposure to IR illumination. Indeed, the prolate band is continuously deformed until  a ``hole'' is created therein, as shown in Fig. \ref{fig:IRvsblue}.

Furthermore, our results suggest that, during relaxation in the dark, the broad distribution of NPs, created by IR light, acts as sink for atoms evaporated from the unstable spheroids. This process slows down the decay of the IR band respect to the nIR one. These facts confirm the hypotheses of a geometry rearrangement of the IR population and of a transfer of atoms from the nIR population to the IR one.

At the same time, the lack of bulk and surface diffusion during the IR illumination causes both the accumulation of larger metallic islands and the survival of random atomic layers which ``bury'' weak adsorbing sites and  potential nucleation centres for the formation of spheroids. As a matter of fact, this process accelerates the decay of the nIR peak.

These observations are consistent with the results reported in the previous section. In fact, IR light, even at low intensity, modifies the substrate available for atoms diffusing along the nanopores. Consequently, the adsorbate-substrate interaction is changed and therefore the system evolves towards a new, long-lasting metastable equilibrium.

\section{Conclusions}
The effect of visible and infrared light on the complex system composed by vapour-phase K atoms, adatoms and metallic aggregates, confined in glass nanopores, has been investigated in this paper.
 
Visible light, which enhances the atomic mobility, produces a plasmon peak in the nIR region, due to nanometric K spheroids \cite{marmugi2014} and a secondary well-defined Gaussian peak in the IR spectrum. This absorption band is attributed to the formation of larger K aggregates, with a broad distribution of sizes and geometries. Such structures self-assemble as a consequence of the interplay between atomic photodesorption, diffusion and NPs' formation and evaporation. Population transfer between the nIR spheroids and the IR K aggregates can be inferred from the evolution of the absorbance spectra. Nevertheless, the number and size of nucleation centres distributed onto the substrate is not significantly affected.

In the case of infrared illumination, instead, we do not observe a significant increase of the desorption efficiency, but we find evidences that the substrate at the pore surface undergoes structural changes, which alter the interaction between the atoms and the surrounding environment. Indeed, infrared light produces a rearrangement of the IR population and the formation of broader metallic islands at the pore surface. Such structures reduce the atomic mobility, as well as the availability of isolated surface defects for cluster nucleation. This implies that the equilibrium condition within the glass matrix is shifted and the system evolves towards a different metastable state.
 
Our results confirm that light can be used as a powerful tool to control the atomic adsorption/desorption and nucleation processes at the nanoscale. Furthermore, under dimensional confinement, exposure to light can also locally modify the substrate structure. Therefore, the substrate-adsorbate interaction can be directly influenced by light, even without an increase of the atomic diffusion through the pores. This last effect extends the processes which can be optically controlled  in nano-sized confined systems, paving the path towards light structuring of metal-composite materials.

\section*{Acknowledgments}
The authors acknowledge the cell manufacturing by M. Badalassi (CNR) and thank C. Stanghini and L. Stiaccini (University of Siena) for their technical assistance. L. M. is currently supported by a Foundation ``Angelo Della Riccia'' scholarship and by the UCL Quantum Science and Technology Institute UCLQ Fellowship.

\section*{References}
\bibliographystyle{iopart-num}
\bibliography{IRbib}

\end{document}